\documentclass{aa}
\usepackage{graphicx}
\usepackage{natbib}
\usepackage{longtable}

\bibpunct{(}{)}{;}{a}{}{,} 

\begin{document}

\title{Magnetic activity of the solar-like star HD~140538}

\author{M. Mittag\inst{1}\and J. H. M. M. Schmitt\inst{1}\and T. S. Metcalfe\inst{2,3}\and A. Hempelmann\inst{1}\and K.-P. Schr\"oder\inst{4,5}}
\institute{Hamburger Sternwarte, Universit\"at Hamburg, Gojenbergsweg
  112, 21029 Hamburg, Germany\\
           \email{mmittag@hs.uni-hamburg.de}
           \and
           Space Science Institute, 4750 Walnut Street, Suite 205, Boulder, CO 80301, USA
           \and
           Max-Planck-Institut f\"ur Sonnensystemforschung, Justus-von-Liebig-Weg 3, 
           D-37077, G\"ottingen, Germany
           \and
           Department of Astronomy, University of Guanajuato, Mexico
           \and Sterrewacht Leiden, University of Leiden, Netherlands}
\date{Received \dots; accepted \dots}

\abstract{
The periods of rotation and activity cycles are among the most important properties of the
magnetic dynamo thought to be operating in late-type, main-sequence stars. 
In this paper, we present a S$_{\rm{MWO}}$-index time series composed from different data sources
for the solar-like star HD~140538 and derive a period of 3.88$\pm$0.02 yr for its activity 
cycle. Furthermore, we analyse the high-cadence, seasonal S$_{\rm{MWO}}$ data taken with the 
TIGRE telescope and find a rotational period of 20.71$\pm$0.32 days. In addition, we estimate 
the stellar age of HD~140538 as 3.7 Gyrs via a matching evolutionary track. This is 
slightly older than the ages obtained from gyrochronology based on the above rotation period,
as well as the activity-age relation. These results, together with its stellar parameters that are
very similar to a younger Sun, make HD~140538 a relevant case study for our understanding of solar activity 
and its evolution with time.}


\titlerunning{Magnetic activity of solar-like star HD~140538}

\maketitle

\section{Introduction}

By diligently counting the number of sunspots on the surface of the Sun, \cite{schwabe1844} 
discovered the eleven-year solar activity cycle, which has hitherto been observed in many 
observational indicators
over almost the entire electromagnetic spectrum.   Since stars cannot be spatially resolved,
Schwabe's technique cannot be applied, and other methods must be used, for example, monitoring
the cores of the Ca II H\&K lines. \citet{eberhard1913} were the first to note, on accidentally 
overexposed plates, ``extraordinarily bright, sharp emission lines in the middle''
\citep[][p.292, ll. 25-26]{eberhard1913}
of the Ca II H\&K lines -- among others -- in the star $\sigma$ Gem, an active RS~CVn binary system,
and interpreted these reversals as ``The same kind of eruptive activity that appears in sun-spots, 
flocculi, and prominences'' \citep[][p.294, ll. 33-34]{eberhard1913}  known from the Sun.
It is remarkable that \citet{eberhard1913}
already considered using the core emission as cycle diagnostics and wrote that ``It remains to be 
shown whether the emission lines of the star have a possible variation in intensity analogous 
to the sun-spot period.'' \citep[][p.295, ll. 3-5]{eberhard1913}.

This idea was taken up in 1966, when O.C. Wilson started the famous Mount Wilson project with a
specially designed HK photometer \citep{wilson1978,Vaughan1978}. Over more than three decades
a systematic search for stellar activity cycles based on the so-called S-index, which measures
the strength of Ca II H\&K emission lines, was carried out. \citet{b95} present
the Mount Wilson project results for 112 stars (including the Sun), with 46 stars showing
evidence of an activity cycle. 

After the official end of the Mount Wilson programme, other programmes continued
Ca II H\&K monitoring with spectroscopic means. Lowell Observatory \citep{Hall1995,Hall2007}
used the  Solar-Stellar Spectrograph (SSS) to this end, but their observations do not cover 
all objects listed in \citet{b95}.  In the last twenty years, other Ca II H\&K monitoring programmes
were carried out, often in the context of radial velocity searches for extrasolar planets, where
S-indices can be derived as a by-product. Catalogues of S-indices are published by see e. g.,
\citet{Henry1996, Gray2003, Isaacson2010ApJ}.

At Hamburg Observatory we continued Ca II H\&K monitoring in 2013, including all
stars listed in \citet{b95}, using our robotic spectroscopy telescope TIGRE (Telescopio
  Internacional de Guanajuato, Rob\'otico-Espectrosc\'opico)
\citep{Schmitt2014AN335787S}, 
located at La Luz Observatory in central Mexico.  Stars listed in \citet{Hall2007} are also
part of the TIGRE activity survey target list to obtain an overlap between the Ca II H\&K programmes 
at Lowell and Hamburg Observatories.  To obtain a longer time span and to increase data 
density, the Ca II H\&K measurements from different sources can be combined, and in this
fashion \citet{Metcalfe2013} found two activity cycles in $\epsilon$ Eri,
which were not listed as cyclic by \citet{b95}.

The wider scope of this research is to study solar magnetic activity by means of younger and 
older, very sun-like stars. Since the sample of such stars with known activity cycles 
and rotation periods is still quite small, any new addition to it is very valuable. The relation 
between the period of an activity cycle and the rotation is particularly relevant in this context, 
because it provides important information about the magnetic dynamo thought to be operating in 
the Sun and solar-like stars \citep{Metcalfe+vanSaders2017}. 
Therefore, the search of new activity cycles is an important task to better 
\begin{figure*}[!t]
\centering
\includegraphics[scale=0.6, angle=90]{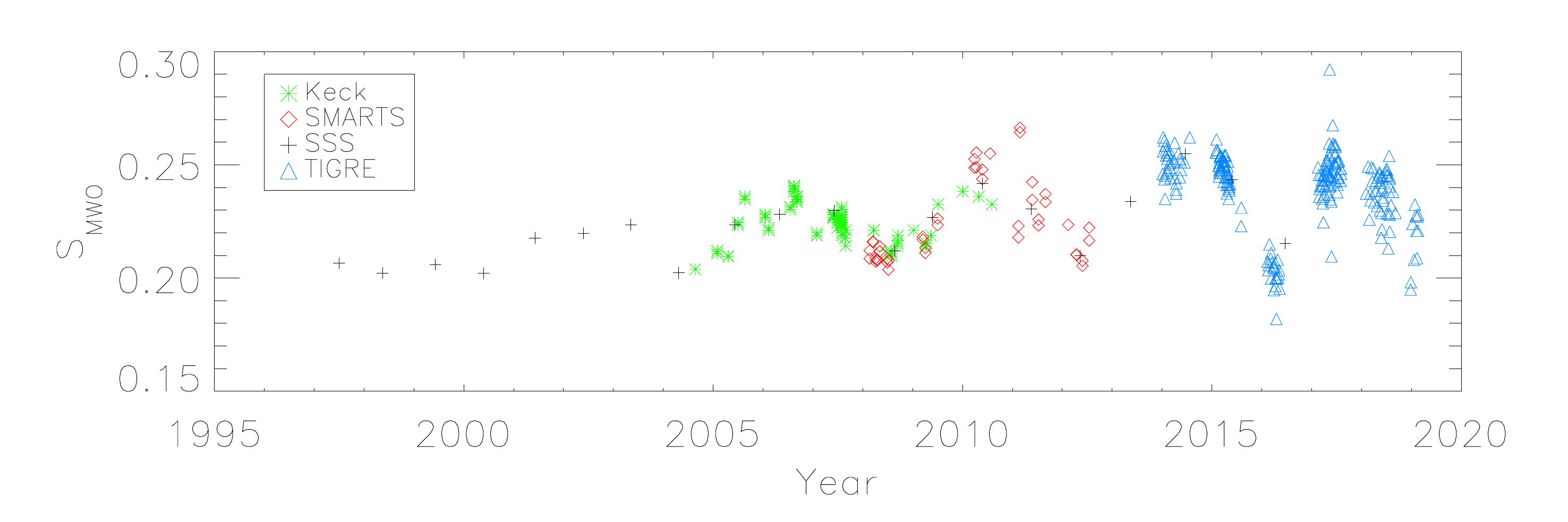}
\caption{ Combined S$_{\rm{MWO}}$ time series. The individual time series are colour-coded and
  labelled with different symbols (Keck: green asterisks, SMARTS: red diamonds,
  SSS: black crosses,  and TIGRE: blue triangles). }
\label{combine_s-index_time_eries}
\end{figure*}
understand and characterise stellar dynamos.  In this paper, we present a combined Ca II H\&K time series of the
solar-like star HD~140538 and the results of our cycle analysis.  We also report a measurement 
of the rotation period of HD~140538, based on high-cadence seasonal data taken with the 
TIGRE telescope, and then estimate the age of HD~140538, based on the independent methods 
of evolutionary tracks, gyrochronology, and activity.
 
\section{Physical parameters of HD~140538}

HD~140538 (= $\psi$ Ser) is a bright (5.86 mag) G2.5 main-sequence star
with a rotational velocity $v sin(i$) of 1.58$\pm$0.13 km/s \citep{dosSanto2016}
at a distance of 14.77$\pm$0.01~pc \citep{Gaia2018yCat.13450G}
and a colour index of B$-$V~=~0.684$\pm$0.002~mag 
\citep{HIPPARCOS1997ESA}. From this brightness and distance an absolute visual magnitude of 5.013$\pm$0.002 mag can be deduced. 
Using the PASTEL catalogue (version 2016-05-02; \cite{Soubiran2016A&A}),
we computed the median of the listed log(g), [Fe/H], and 
T$_{eff}$ values without double recorded values. Furthermore, standard
deviations of each median are computed from the median absolute deviation and
assumed these values as the uncertianty of the parameters.
We obtain for log(g) a median of 4.46$\pm$0.06, for [Fe/H] a median of 0.05$\pm$0.02,
and for T$_{eff}$ a median of 5683$\pm$15~K. A comparison of these values with the solar values shows 
HD~140538 to be slightly cooler and its metallicity to be slightly higher, but overall 
the stellar parameters of HD~140538 are very close to those for the
Sun, suggesting that HD~140538 is (close to being) a solar twin. On the other hand, 
\citet{Mahdi2016A&A} provide a study with a list of solar twin candidates that includes
HD~140538, but they find differences in abundance and age compared to the Sun, and
consequently do not include HD~140538 in their final list of solar twin stars. 

\section{Monitoring chromospheric activity by S-index data}

To study the long-term chromospheric activity of HD~140538, we used the S$_{\rm{MWO}}$ 
values from four different data sources and combine all the available data into a 
single data set with a total of 
438 individual S$_{\rm{MWO}}$ measurements over the time period of 1997 to now. 
Our time series is displayed in Fig.~\ref{combine_s-index_time_eries}. In the following we briefly describe
the individual data sets.

\subsection{Keck}

We used the S$_{\rm{MWO}}$-values from the catalogue published by \cite{Isaacson2010ApJ}, who
provide S$_{\rm{MWO}}$ time series for more than 2600 stars. These S-values were obtained
in the context of the California Planet Search (CPS) programme with telescopes at the Keck and 
Lick Observatories. For our combined time series, we use only the S-index values obtained from 
the spectra taken at the Keck Observatory, since the Lick data show a rather large scatter.  
We use a total of 136 S$_{\rm{MWO}}$-values taken between  2004 to 2012.

\subsection{SMARTS}

To bridge the gap between the measurements from Keck and TIGRE, we used
observations from the SMARTS southern HK project \citep{Metcalfe2010,
Metcalfe2013}. These data include 45 low-resolution (R$\sim$2500) spectra
obtained at 23 epochs between February 2008 and July 2012 using the {\it 
RC~Spec} instrument on the 1.5-m telescope at Cerro Tololo Interamerican
Observatory. Bias and flat-field corrections were applied to the 180~s    
integrations, and the wavelength was calibrated using standard IRAF
routines. S$_{\rm{MWO}}$ values were extracted from the reduced spectra    
following \cite{Duncan1991}, placing the instrumental measurements onto
the Mount Wilson scale from contemporaneous observations of 26 stars from  
the SSS.

\subsection{Solar-Stellar Spectrograph}

We use S$_{\rm{MWO}}$ data from the
SSS published by \citet{Radick2018ApJ}. It should be noted that
these data are seasonal or rather yearly S-values covering
a time span of 20 years from 1995 to 2016, constituting the longest
individual time series in our data set. The total number of individual 
S-index values in a single season 
varies between 8 values (in 2008) and 38 values (in 2009), with an
average of 24 S-index values per season.   

\subsection{TIGRE}

We finally used our own TIGRE S-indices estimated from spectra taken with the fully 
robotic TIGRE telescope. The TIGRE telescope has an aperture of 1.2 m and its 
only instrument is the two spectral channel fibre-fed \'echelle spectrograph
HEROS (Heidelberg Extended Range Optical Spectrograph) with a spectral
resolution of R$\approx$20000 covering the wavelength range between
3800 to 8800 \AA\ with a small gap between the two spectral channels;
a more detailed description of the TIGRE telescope is given by \cite{Schmitt2014AN335787S}. 
The estimation of the instrumental TIGRE S-index and its transformation onto
the Mount Wilson scale is described by \cite{Mittag2016}. 
The total number of TIGRE spectra used is 238, covering the time span between 2014 and 2019. 

\subsection{Combined S-index time series}

When combining the four individual data sets into a single data set, we had to consider 
possible offsets of individual data sets caused
by small differences in the calibration into the Mount Wilson scale. 
This expected misalignment was visible and had to be corrected.
For that, we used the SSS data as a reference and re-scaled the other three data sets to the scale of
the SSS data. To re-scale the time series, the mean of the S-values of the time series
in the time overlap regions were computed, and the ratio between the mean value of the
SSS data and the other time series was used as scaling factor.
In Table \ref{tab1}, we list the number of S$_{\rm{MWO}}$ values,
the mean re-scaled S$_{\rm{MWO}}$, the standard deviation, the log R$_{\rm{HK}}^{+}$ \citep{mittag2013A&A549A117M},
and the scale factor for each individual time series. 
The combined and corrected time series is displayed in Fig.~\ref{combine_s-index_time_eries}.
The individual time series are colour-coded and labelled with different symbols.  

\section{Magnetic activity: cycle, rotation, and age}

In this section we present the results of our investigation of the stellar activity 
based on the combined S$_{\rm{MWO}}$ time series of HD~140538, and also
discuss the general chromospheric and coronal activity level of HD~140538.
We performed a Lomb-Scargle analysis of the S$_{\rm{MWO}}$  time series to
estimate the activity cycle, and discuss a possible rotational period found in the TIGRE data.

\subsection{Chromosphere}
\label{chrom}
HD~140538 is listed in the Washington Double Star Catalogue \citep{Mason2001AJ}
with five individual components. However, these are much fainter than the main object
so that for the S-index, the influence of these additional components is negligible.
The individual S$_{\rm{MWO}}$ time series all have a different number of data points. 
Therefore, we calculated the mean S$_{\rm{MWO}}$
values for each individual corrected time series and list these values in Table \ref{tab1}.
The values in Table \ref{tab1} yield a mean S$_{\rm{MWO}}$ of 0.228 with a standard deviation of 0.008.
Given that HD~140538 is close to being a solar twin, we can directly compare 
the activity level of HD~140538 with the chromospheric activity level of the Sun.  Using a mean 
solar S$_{\rm{MWO}}$-value of 0.1694$\pm0.0005$ \citep{Egeland2017ApJ}, we
estimate a $\approx$ 35$\%$ higher chromospheric activity level of HD~140538 compared to the Sun. 

\begin{table}[!t]    
\caption{Data sets, number of S$_{\rm{MWO}}$ values, mean re-scaled S$_{\rm{MWO}}$, standard deviation ($\sigma$), corresponding
    log R$_{\rm{HK}}^{+}$ for each time series and re-scaling factor}    
\label{tab1}    
\begin{center}    
\begin{small}    
\begin{tabular}{cccccc}    
\hline    
\hline    
\noalign{\smallskip}    
Data set & No. & $\overline{\rm{S}_{\rm{MWO}}}$ & $\sigma$ & log R$_{\rm{HK}}^{+}$ & Scale factor \\    
\hline    
\noalign{\smallskip}    
Keck   & 136 & 0.226 & 0.008 & -4.67 & 1.15\\
SMARTS & 45  & 0.224 & 0.018 & -4.68 & 1.11\\
SSS    & 20  & 0.221 & 0.015 & -4.69 & 1.00 \\
TIGRE  & 238 & 0.240 & 0.017 & -4.61 & 1.04\\
\hline    
\end{tabular}
\tablefoot{The scale factor is that required to re-scale the time series to the SSS S-index time series.}
\end{small}    
\end{center}    
\end{table}

\subsection{X-ray emission}

HD~140538 has apparently never been the target of a dedicated X-ray observation according to the archives of the XMM-Newton and ROSAT missions. 
However, X-ray emission from HD~140538 has been detected in the context
of the ROSAT all-sky survey (RASS).  Inspecting the catalogue of 
RASS X-ray sources by \cite{boller2016}, we identify the X-ray source
2RXS~J154402.0+023054 with HD~140538. \cite{boller2016} report 2RXS~J154402.0+023054
with a count rate of 0.15 $\pm$ 0.02 cts/s. The positional coincidence and the softness
of the recorded spectrum make this identification essentially certain; no indications
of time variability are found in the RASS data.

Using the count-rate-to-flux conversion by \cite{schmitt1995} we find an apparent soft 
X-ray flux of 7.3 $\times$ 10$^{-13}$ erg/cm$^2$/s, which translates into an X-ray luminosity of
1.9 $\times$ 10$^{28}$ erg/s, using a distance of 14.77~pc \citep{Gaia2018yCat.13450G}.
Thus, although very similar to the Sun in a variety of aspects, HD~140538 exceeds
the X-ray output of the active Sun by at least an order of magnitude, which is in line with its
likewise enhanced chromospheric activity.

\begin{figure}  
\centering
\includegraphics[scale=0.3, angle=90]{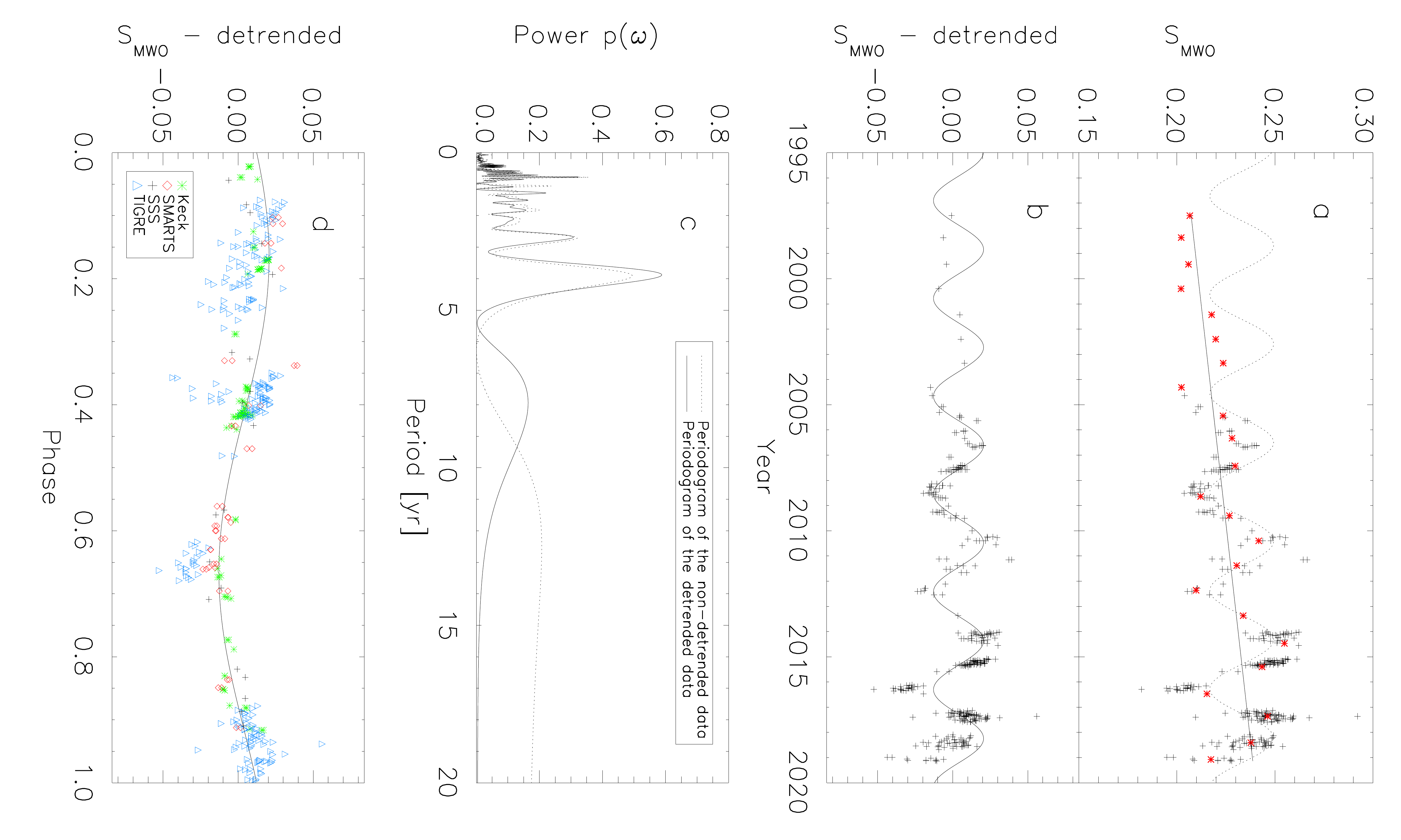}
\caption{Panel a: Combined S$_{\rm{MWO}}$ time series is displayed with black crosses. 
The red asterisks  show the seasonal data from SSS and TIGRE data and the solid line the 
result of the linear trend fit. The dotted line depicts the sinusoidal fit of the of
the Lomb-Scargle analysis of the non-detrended data with a period of 3.91 yrs.
Panel b: Detrended S$_{\rm{MWO}}$ time series. The solid line represents the sinusoidal fit with the period of 3.88 years
found in the GLS analysis. Panel: Periodograms
of  GLS (generalised Lomb-Scargle) analysis of detrended and non-detrended data. Panel d: Phase-folded detrended S$_{\rm{MWO}}$ 
time series shown with sinusoidal fit. Here, the S$_{\rm{MWO}}$ of the 
individual time series are colour-coded and labelled (Keck: green asterisks,
SMARTS: red diamonds, SSS: black crosses,  and TIGRE: blue triangles).}
\label{cycle_result_time_series}
\end{figure}

\subsection{Activity cycle}

A visual inspection of the S-index time series of HD~140538
shows a clear  approximately four-year periodic variation, which was already noted by
\cite{hall2007AJ}.  An extended time series of SSS S-index values
is shown by \cite{Hall2009AJ} and \citet{Radick2018ApJ}.  Here we perform a
rigorous period analysis based on the Lomb-Scargle periodogram.

For this analysis, we used the generalised Lomb-Scargle (GLS) formalism
by \citet{Zechmeister2009}.
This formalism is based on a $\chi^{2}$ minimisation with the model
$y(t)=a\sin(\omega t)+b\cos(\omega t)+c$, where c is
a constant. Therefore, the peak of the periodogram provides not only the probability of the period, it
is also a sign of the quality of the $\chi^{2}$ fit by the given period. To estimate the error
of the period, we used the error equation for the period from \citet[][Eq. 3]{b95}.

The SSS S-index values published by \citet{Radick2018ApJ} show a slightly increasing trend, 
which is also visible in the combined S-index time series. This may be an indication of a
long-term activity cycle.
However, we performed a Lomb-Scargle analysis without the removal of this trend to test
its influence on the period estimation. The periodogram is shown as a dotted
line in Fig. \ref{cycle_result_time_series}, panel c, and we find a clear peak at the period
of 3.91$\pm$0.02 years. The peak height is 0.495, which corresponds to a
formal FAP (false alarm probability) of 4$\times10^{-62}$, in other
words, this periodicity is very certain. The sinusoidal fit is shown in
Fig. \ref{cycle_result_time_series}, panel a as a dotted line. It can be seen
that the fit is systematically higher for the data before 2005. In the next step, we removed this trend
from the S-index data because the trend is not considered in the GLS formalism
and could have an influence on the period estimation. 

For the trend estimation, we used only the seasonal SSS S-index values
because these data are the most homogeneous and form longest set of our four data sets. For the 
seasons from 2017 to 2019, we extended the SSS data with the seasonal TIGRE S-index values from 2017.
In panel (a) of Fig. \ref{cycle_result_time_series}, these seasonal values are shown as red 
asterisks with all S-index values as black crosses.
For the trend, we use a linear approach and the result of the fit is plotted in 
Fig. \ref{cycle_result_time_series}, panel (a) as a
solid line. After removing the trend (see Fig.~\ref{cycle_result_time_series}, 
panel (b)), we perform a Lomb-Scargle analysis, using GLS formalism. 
The periodogram of this analysis is shown in Fig.~\ref{cycle_result_time_series}, 
panel (c). We find a clear signal with a period of 3.88$\pm$0.02 years, a peak
height of 0.588, and a corresponding 
formal FAP of 2$\times10^{-81}$,
which is higher than without the trend removal. This is also an indication for a trend in the
data because the GLS model fits better after the removal of the trend. This is visible in
Fig. \ref{cycle_result_time_series} when comparing both fits. Therefore, we prefer the period obtained
after the trend removal. On the other hand, both periods
are equal within the error and this shows that de-trending has only a very small effect on
the period estimation. The possible reason is that the majority of data points are taken after 2005,
so that the influence of the trend on the period estimation and the data before 2005 is small. 

Independent of the de-trending, the period of the last cycle of this time series is shorter
than in the other cycles, which we roughly estimate at a period of $\approx$3.4 years. Therefore,
the data do not follow the sinusoidal fit with
the 3.88-year period. To test the influence of the last shorter cycle on the 3.88-year
period, we estimated the period without this cycle and obtained a period of 3.99$\pm$0.02 years
with a formal FAP of 1$\times10^{-88}$. However, period variations of $\approx$10$\%$ are known to
occur from the well-studied solar cycle, so this may already explain the discrepancy. 

In the plot of the residuals (see Fig.~\ref{cycle_result_time_series} panel (b)) a negative trend
starting in about 2015 can be observed. This could indicate that the cycle maximum of the possibly
longer term cycle has passed and the activity is decreasing slowly
to the next minimum of the longer term cycle. To verify this, we need longer observations. However,
if this true, half of the long-term cycle can be seen under the additional assumption that at the
beginning of the time series, the star was in the minimum
of this long-term cycle. Therefore, a period of this long-term cycle can be roughly estimated at around 30 years. 

\subsection{Rotation period}
\label{rotation}
\begin{figure}  
\centering
\includegraphics[scale=0.3]{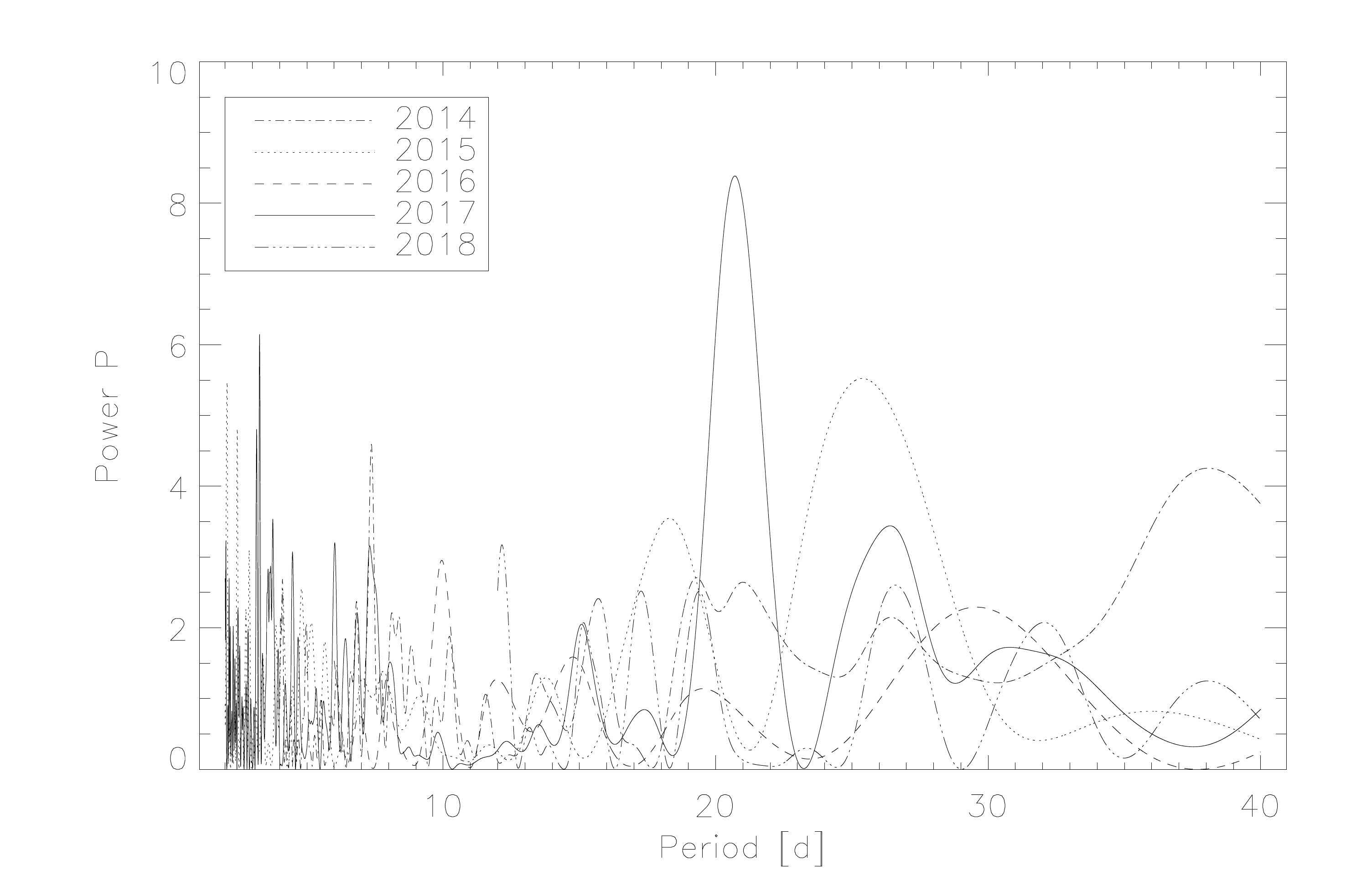}
\caption{Periodograms of GLS analysis for each TIGRE observation season. The years are shown in the following manner: 2014
by a dashed-single-dotted 
line; 2015 a dotted line; 2016 a dashed line; 2017 a solid line; 2018 a dashed-triple-dotted
line.}
\label{seasonal_periodogram}
\centering
\includegraphics[scale=0.3]{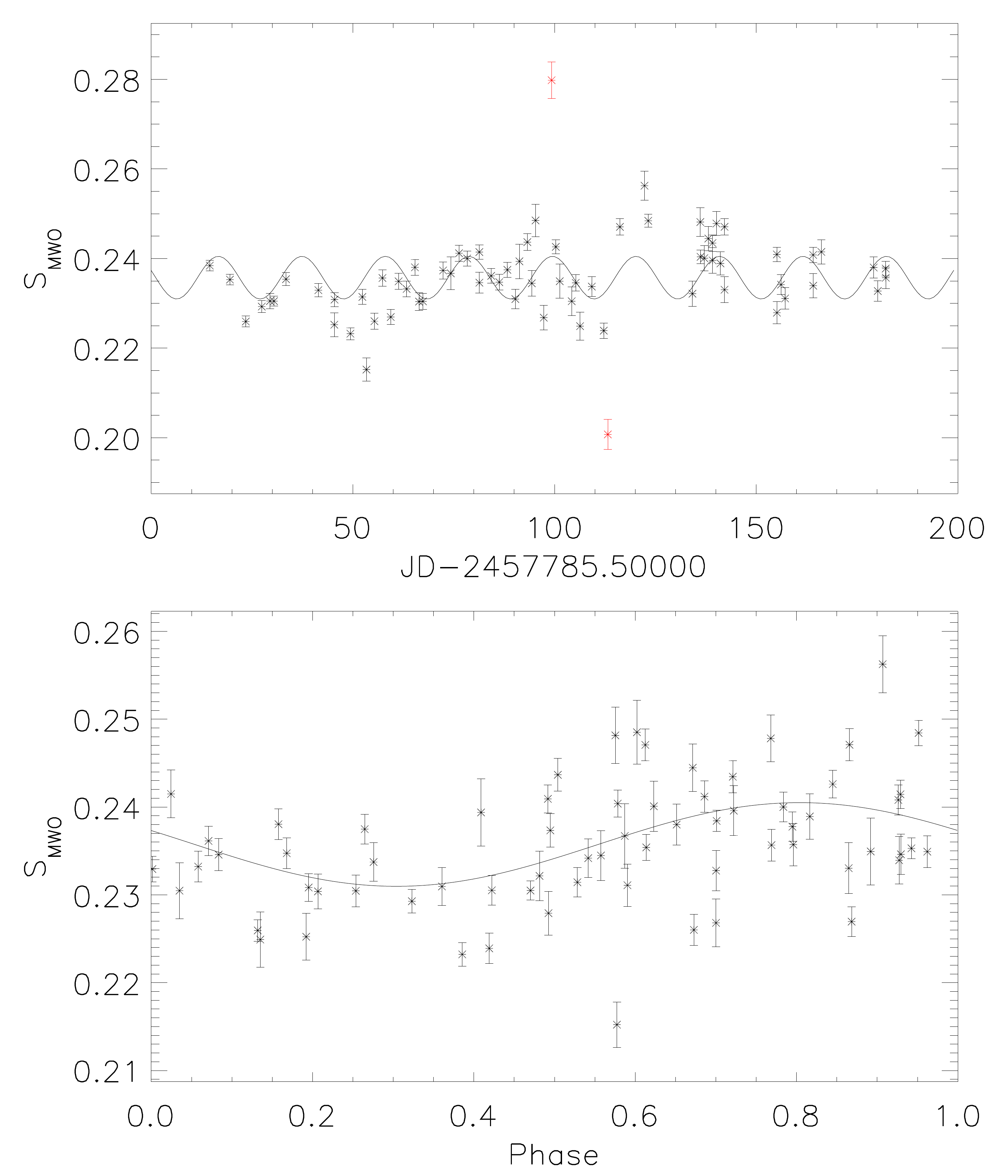}
\caption{Upper panel: TIGRE S$_{\rm{MWO}}$-index values of  2017 season with two outliers plotted
  as red data points. The solid line shows the sinusoidal fit with the 20.71 days rotational period.
  Lower panel: Phase-folded S$_{\rm{MWO}}$-index time series shown with rotational period of 20.71 days. The solid line depicts the best sinusoidal fit.}
\label{rotation_period}
\end{figure}

The rotation period of a star is a key parameter to characterise its stellar activity.
Our TIGRE measurements are sufficiently densely sampled to allow meaningful period measurements.
In Fig.~\ref{seasonal_periodogram} we show the GLS periodograms ($P = \frac{N-1}{2}p(\omega)$), 
including the noise level) for each observational
season, the individual seasons are shown with different line styles. 
Somewhat surprisingly, only in the season of 2017 could we find a significant rotational period 
at 20.71$\pm$0.32 days, (see Fig.~\ref{seasonal_periodogram}). 
The formal FAP of this peak is 0.007 with a significance of 99.3$\%$.
The upper panel of Fig.~\ref{rotation_period} shows the TIGRE S$_{\rm{MWO}}$ values
recorded in the 2017 season, where the two values plotted in red are treated as outliers and
thus ignored in the GLS analysis. The solid line depicts the sinusoidal fit with the determined 
rotational period. In addition, the phased-folded TIGRE S$_{\rm{MWO}}$ values for this
2017 season are shown with the best sinusoidal fit as a solid line in the lower panel of 
Fig.~\ref{rotation_period}.

Furthermore, we were able to use the estimated mean S$_{\rm{MWO}}$-index of 0.228 to calculate the 
rotational period from the activity-rotation relation.
First, we converted the mean S$_{\rm{MWO}}$-index of 0.228 into the R$_{\rm{HK}}^{+}$ value 
using the conversion developed by \citet{mittag2013A&A549A117M} and 
obtain R$_{\rm{HK}}^{+}10^{5}$ = 2.18$\pm$0.22. From that, we estimated the Rossby 
number of 0.5 from the activity-rotation relation developed by \citet{mittag2018}. With this 
Rossby number and the convective turnover time used in \citet{mittag2018}, we calculated a rotational
period of 21.3$^{+1.3}_{-1.2}$ days for HD~140538 in good agreement with our TIGRE measurements; 
we note that \citet{Isaacson2010ApJ} also arrived a rotational period of about 20 days.
The comparison with the calculated periods shows that the rotational period obtained
from the TIGRE S$_{\rm{MWO}}$-index time series are reasonable although the significance of the 
results is slightly lower than the formal 3$\sigma$ level.

\subsection{Stellar age}
\label{age}
\begin{figure}  
\centering
\includegraphics[scale=0.3]{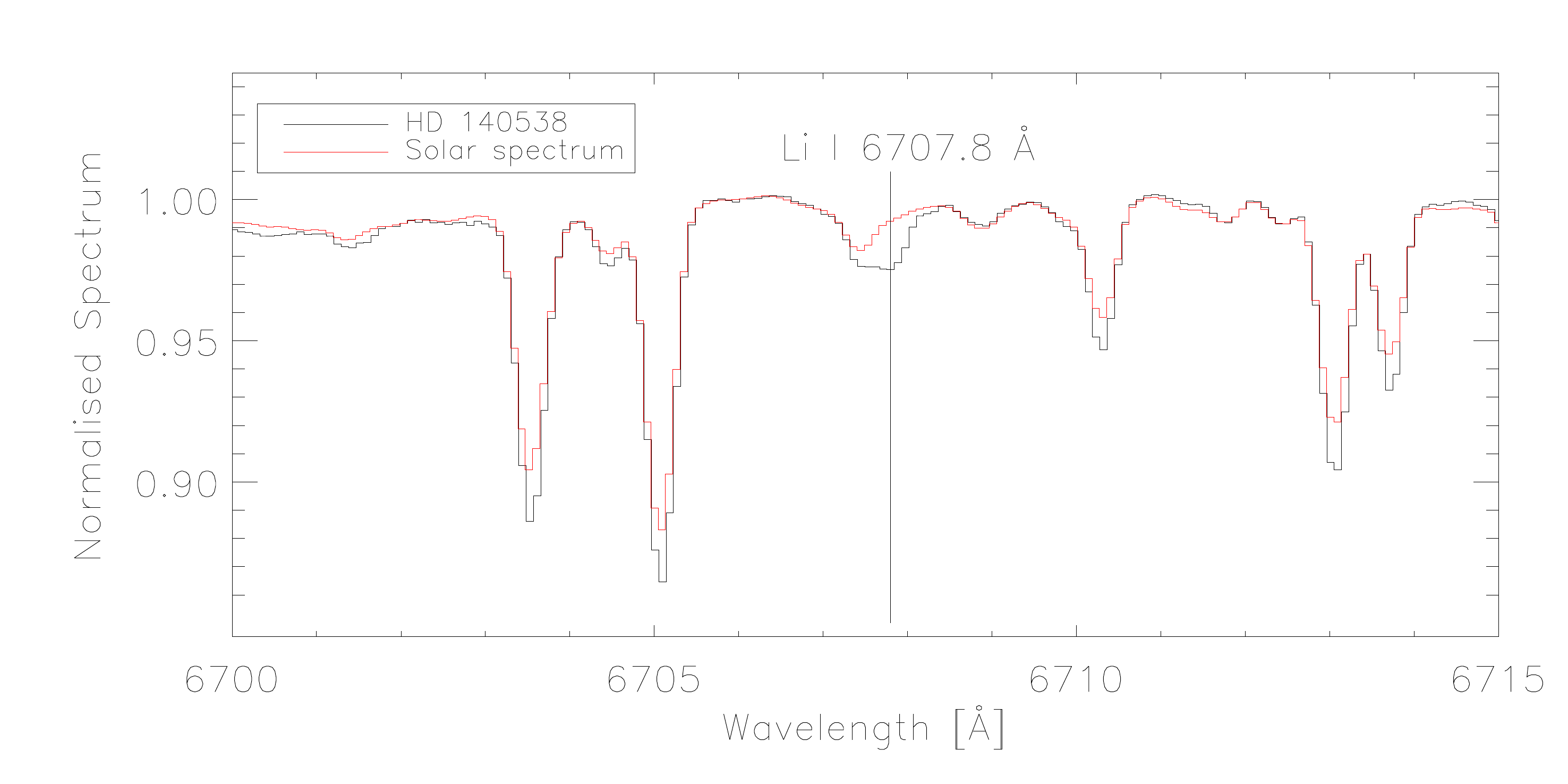}
\caption{A comparison of the co-added spectrum of HD~140538 (black line) and co-added solar spectrum (red line).}
\label{lithium_line}
\centering
\includegraphics[scale=0.3]{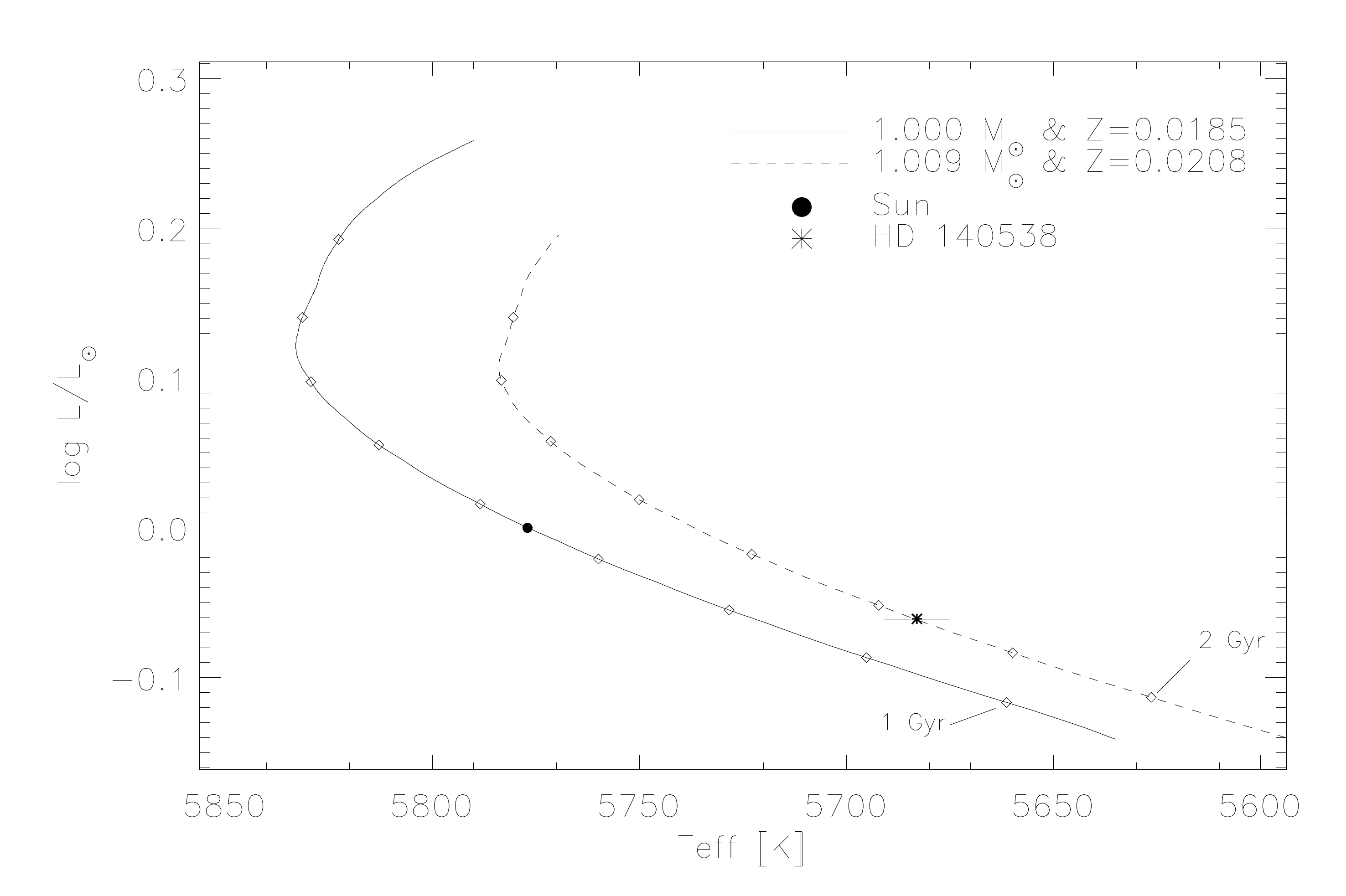}
\caption{Luminosity vs. effective temperature of the stars.   
  The solid line shows the evolutionary track for 1.00$M_{\odot}$ with Z=0.0185 and
  the dashed line for 1.009$M_{\odot}$ with Z=0.0208. The diamonds on the evolutionary   
  track mark time steps of 1 Gyr, with the first marked time step at 1 Gyr and at 2 Gyr, respectively.}
\label{evol_tracks}
\end{figure}

In the context of gyrochronology, stellar rotation is directly related to 
stellar age \citep{Barnes2007ApJ}, and, in a wider context, 
to the evolutionary stage of the star (see, among a vast literature, \citet{schroeder2013} 
and references therein). Despite stellar age being such an important stellar parameter, it 
is rather difficult to determine in absolute terms.  Here we use different methods
to assess the likely age of HD~140538.
 
We first inspect our TIGRE spectra for the presence of
the lithium line at 6707.8 \AA. In Fig.~\ref{lithium_line} we show the co-added TIGRE spectra of
HD~140538 (black line) and the Sun (measured from Moon light spectra, indicated by the red line),
where the clear presence of lithium absorption in contrast to the Sun is observable.  While actual 
lithium `clocks' are quite controversial, we conclude that HD~140538 is younger than the Sun, 
as expected from its faster rotation.

We next used evolutionary tracks to quantitatively determine the age of HD~140538. 
To that end we used the evolutionary tracks with the Cambridge Stellar Evolution
Code\footnote[1]{http://www.ast.cam.ac.uk/$^\sim$stars/} \citep{Pols1997MNRAS.289.869P}, 
which is an updated version of the evolution code written  by \cite{Eggleton1971MNRAS.151..351E}. 
The code requires values for stellar mass and metallicity as input parameters.  As a first step 
it is necessary to fine-calibrate the code with the solar values.
Here we used the empirical calibration from \citet{Mittag2016} and plot the resulting evolutionary 
track (as a solid line), as well as the position of the
Sun (as a filled circle) in the HRD. This model is calculated with Z=0.0185 
([Fe/H]=0.00) and suggests a solar age of around 4.6 Gyr,
which is consistent with the age obtained by \cite{2010NatGe...3..637B}.

Using the absolute visual magnitude of 5.013$\pm$0.002~mag and
the bolometic correction (BC) of -0.12~mag calculated from \citet[][Eq. 10.9]{Gray2005oasp.book},
we find a luminosity of $\log L/L_{\odot}$~=~$-$0.061$\pm$0.001 for HD~140538 with M$_{\rm{bol},\odot}$=4.74~mag
\citep{Cox2000}.
With this value and an effective temperature of $T_{\rm{eff}}$ of 
5683$\pm$15~K, the position of HD~140538 in the HRD is well defined and shown as 
an asterisk in Fig. \ref{evol_tracks}; note
that the parallax error of HD~140538 is very small and thus negligible for the age estimation.
The uncertainty in $T_{\rm{eff}}$ also influences the stellar age estimate since a variation
in stellar mass and the uncertainty in metallicity causes a shift in the effective temperature.
To consider these variations, we estimated the stellar age for $T_{\rm{eff}}$ $\pm$15 K with
fixed metallicity by variation of the stellar mass. Furthermore, we estimated the stellar age and
mass of the best-matching
tracks for Z=0.0208 $\pm$0.001, which is the corresponding the error range of the metallicity.
The evolutionary track for HD~140538 is then calculated with M = 1.009 M$_{\odot}$ and Z=0.0208 and 
depicted as a dashed line in Fig.~\ref{evol_tracks}.  
This evolutionary track suggests 3.7$\pm$1.6 Gyr as the likely
stellar age of HD~140538, and we estimate a stellar mass of M = 1.01$\pm$0.02 M$_{\odot}$. 
These results show that HD~140538 is younger than our Sun with a mass that is possibly slightly higher.

In addition, we estimated the age of HD~140538 from the mean activity of 
HD~140538 and the rotation period of this star derived in Sect.~\ref{chrom} and \ref{rotation}.
To obtain the age from the rotation, we used the method of gyrochronology
following \citet{Barnes2007ApJ} and calculated an age of 2.4$\pm$0.3~Gyr. From the activity, 
we calculated an age of 2.2$\pm$0.3~Gyr with the age-activity relation derived by 
\citet{Mamajek-Hillenbrand2008ApJ} and a $\log R_{\rm{HK}}^{'}$ of -4.72 using the transformation
equations from S$_{\rm{MWO}}$ into $\log R_{\rm{HK}}^{'}$ from \citet{noyes1984}.
Both ages are consistent with the stellar age to within the error obtained from the evolutionary track.

Finally, we checked the catalogues in the Simbad/VizieR database to find other age estimates. 
\citet{Mamajek-Hillenbrand2008ApJ} list HD~140538 with an age of 3.2 Gyr and a $\log R_{\rm{HK}}^{'}$ of -4.80
taken from \citet{hall2007AJ}, while \citet{Casagrande2011} quote an age of 0.3~Gyr and
\citet{Takeda2007} an age of 7.72 Gyr.  Given the activity properties of  HD~140538, the latter two
age estimates appear quite unlikely, while the age estimated by \citet{Mamajek-Hillenbrand2008ApJ}  is
obviously consistent with our estimates.

\section{Discussion and conclusion}

In this work, we combined the S$_{\rm{MWO}}$ measurements from the 
SSS project, Keck data (CPS), SMARTS
data (southern HK project) and TIGRE data to study the magnetic activity of HD~140538.  
In our long-term time series we find a clear activity cycle with a cycle period
of 3.88$\pm$0.02 yr. In each individual data set, this variation is also clearly visible,
however, the time series also shows a long-term trend that might be an indication for a 
long-term activity cycle. With the derived activity cycle period of 3.88~yr and the 
rotational period of 20.71~day, HD~140538 fits very well in the P$_{\rm{cyc}}$ vs. 
P$_{\rm{rot}}$ distribution for stars on the short, inactive 
branch shown in \citet[][ Fig. 4]{Brandenburg2017}.

The mean S$_{\rm{MWO}}$-index of HD~140538 of 0.228 is clearly larger than the mean
S-value of the Sun with 0.1694 \citep{Egeland2017ApJ}, and the X-ray luminosity obtained with ROSAT shows
HD~140538 to be far more active than our Sun.  In our seasonal TIGRE data, we find a weak periodic 
signal at a period of 20.71 day with a significance
of 99.3$\%$, which we interpret as the rotational period of  HD~140538.
All this evidence points to a star younger than the Sun, and indeed our age estimates 
result in values less than the solar age, independently of whether ages based on
age-activity relations, gyrochonological ages, or evolutionary ages are considered.

Inspecting our long-term time series it appears that in the early 1990s HD~140538 was 
in a minimum of the above long-term cycle. The cycle amplitude also seems to have changed, 
with amplitude variations of the 3.88~yr cycle becoming stronger with time.
Comparable behaviour in the short-term cycle was also observed in the $\epsilon$ Eri time
series published by \citet{Metcalfe2013}:
In $\epsilon$ Eri the periodic variations caused by the short-term cycle were not visible
during the minimum of the long-term cycle, behaviour \citet{Metcalfe2013} assumed to be caused 
by two different dynamos producing the two cycle branches shown in \citet{boehm-vitense2007}.

Since the stellar parameters of HD~140538 are very close to those of the Sun, but 
HD~140538 has three-quarters of the solar age, HD~140538 presents a very good case in the context of the 
Sun-in-time theme. Further S-index monitoring of HD~140538 is necessary to find and measure 
the suspected longer cycle period (expected from the combined S-index time series and
from the long-period branch \citep{Brandenburg2017}) 
from an entire, longer time series. Since the cycle timescales in HD~140538 are three times shorter than 
in the case of the Sun (which is a Schwabe cycle of eleven years, a long-period cycle is possibly the Gleisberg cycle of 
about a century), we expect a period of about thirty years, which requires patience and stamina to
measure.

\begin{acknowledgements}
We thank the referee, Ricky Egeland, for the helpful comments and suggestions.
T.S.M. acknowledges support from a Visiting Fellowship at the Max Planck
Institute for Solar System Research. The authors from 
Hamburg and Guanajuato are grateful for 
financial support from the Conacyt-DFG bilateral project No. 278156,
which enables us to intensify our collaboration by mutual visits. 
This research has made use of the VizieR catalogue access tool, CDS,
Strasbourg, France (DOI: 10.26093/cds/vizier). The original
description of the VizieR service was published in A\&AS 143, 23.
\end{acknowledgements}

\end{document}